# RF thermal and new cold part design studies on TTF-III input coupler for Project-X

PEI Shilun(裴士伦)[1; 1)]  Chris E Adolphsen[2]  LI Zenghai(李增海)[2]  Nikolay A Solyak[3]  Ivan V Gonin[3]

[1]Institute of High Energy Physics, Chinese Academy of Sciences, Beijing 100049, China
[2]SLAC National Accelerator Laboratory, Menlo Park, California 94025, U.S.A.
[3]Fermi National Accelerator Laboratory, Batavia 60510, Illinois, U.S.A.

**Abstract:** RF power coupler is one of the key components in superconducting (SC) linac. It provides RF power to the SC cavity and interconnects different temperature layers (1.8K, 4.2K, 70K and 300K). TTF-III coupler is one of the most promising candidates for the High Energy (HE) linac of Project X, but it cannot meet the average power requirements because of the relatively high temperature rise on the warm inner conductor, some design modifications will be required. In this paper, we describe our simulation studies on the copper coating thickness on the warm inner conductor with RRR value of 10 and 100. Our purpose is to rebalance the dynamic and static loads, and finally lower the temperature rise along the warm inner conductor. In addition, to get stronger coupling, better power handling and less multipacting probability, one new cold part design was proposed using 60mm coaxial line; the corresponding multipacting simulation studies have also been investigated.

## 1  Introduction

Project X is a high intensity proton facility conceived to support a world-leading program in neutrino and flavor physics over the next two decades at Fermilab. The RF coupler requirements for the HE linac are depicted in Table 1 [1]. For safety margin consideration, the coupler should be able to handle ~2.2MW (~10% overhead) peak power during the ~0.2ms filling time and ~550kW (~10% overhead) during the ~1.25-2.5ms flat top with total average power ~15kW [2].

Table 1: RF coupler requirements for the HE linac

| Parameters | 1MW | 2MW upgrade | 4MW Upgrade |
|---|---|---|---|
| Beam energy /GeV | 8 | 8 | 8 |
| Current / mA | 20 | 20 | 20 |
| Repetition rate / Hz | 5 | 10 | 10 |
| Gradient ($\beta$=1) / MV/m | 25 | 25 | 25 |
| $Q_{ext}$ / $10^6$ | 1.25 | 1.25 | 1.25 |
| Filling time / ms | 0.212 | 0.212 | 0.212 |
| $T_{pulse}$ (flat-top) / ms | 1.25 | 1.25 | 2.5 |
| $T_{total}$ / ms | 1.465 | 1.465 | 2.712 |
| $P_{peak}$ (coupler) / kW | 500 | 500 | 500 |
| $P_{average}$ (coupler) / kW | 3.7 | 7.3 | 13.6 |

TTF-III coupler is one of the most promising candidates for the HE linac since it is a proven component of the European XFEL design. It has been demonstrated that TTF-III coupler can handle ~2MW peak power and up to 10kW average power with air cooled central conductor [3]. However, the TTF-III coupler may meet the peak power requirements for all Project X operating scenarios, but cannot meet the average power requirements because of the relatively high temperature rise on the warm side inner conductor, which might result in melting or desquamating of the cooper coating. Some design modifications will be required; one simple way is to increase the copper coating thickness on the warm inner conductor.

It has been known that for the dynamic RF power losses the higher copper RRR is better, but for the static thermal losses lower RRR is better. In order to rebalance the dynamic and static loads, the best compromise for RRR value of copper coating should be found by studying RF-thermal effect at different copper RRR value. For TTF-III coupler, the RRR upper limit was set at 80 to satisfy the requirement of maximum thermal power transmitted at the 4K shield of 0.5W by every coupler [4].

Here for TTF-III like coupler applied to Project X, we studied the RF-thermal effects with different copper coating thickness on the warm inner conductor when the copper RRR values are 10 and 100 respectively. In addition, to facilitate the multipacting problem [2] and get better power handling capability, one new cold part design using 60mm coaxial line was proposed.

## 2. TTF-III coupler

Fig. 1 shows the TTF-III RF input power coupler design [5, 6]. It has 4 fixed temperature layers: outside connection layer to 300K room temperature environment, 70K and 4.2K shield connection layers to cryogenics, and the SC cavity flange at 1.8K temperature layer. Except the coupler cold part antenna is made of pure copper, the other inner and outer conductors are made of stainless steel but coated with copper. TTF-III coupler has 2 ceramic windows (warm and cold) and 3 bellows. Standard TTF-III coupler configuration has 10μm copper coating on both cold and warm outer conductors, but 30μm on warm inner conductor.

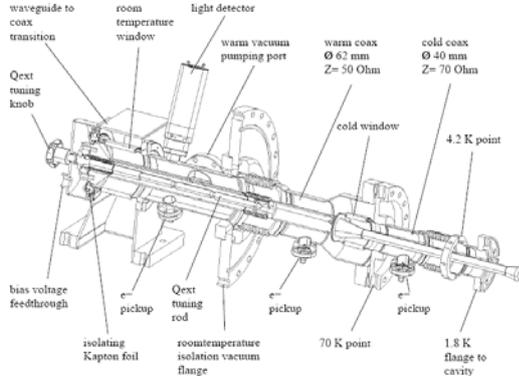

Fig. 1: TTF-III RF input power coupler design

## 3. ANSYS simulations

With the High Frequency and Steady State Thermal solver modules in the multi-physics software package ANSYS [7], numerical RF-thermal coupled finite element analysis (FEA) has been carried out. Currently the ANSYS High Frequency module has the limitation that only 3D elements can be used. To minimize CPU time and memory usage, one axis-symmetric 3D model with 1º azimuth angle was created to perform the analysis. By using one program for all the simulations any problems of transferring loads were eliminated.

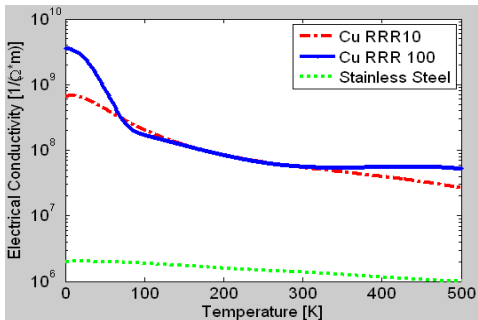

Fig. 2: Electrical conductivities for different temperature

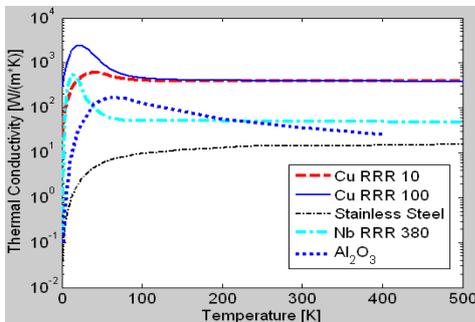

Fig. 3: Thermal conductivities for different temperature

A complete analysis cycle required 5 steps as outlined below. Due to the temperature dependence of material properties shown in Fig. 2 and 3 [6], Step 5 needs to be iterated several times to get a stable thermal solution until the error between two consecutive iterations reaches the specified value. The simulation results in Step 4 serves as the initial condition.

1) The vacuum volume and the copper coating volume were meshed with a common surface interface mesh. The analysis domain volumes were defined with APDL (ANSYS Parametric Design Language) macro language [7]. The common surface mesh created at this step is the key for ease of transfer of the RF wall losses onto the thermal model.

2) The copper coating volume and the rest metallic volumes were meshed with SOLID90 20-Node thermal solid element, but assigned with different materials. The pure copper antenna was also modeled with two separate volumes: copper coating and copper metallic volumes.

3) With HF120 high frequency brick solid element, harmonic analysis was performed in the vacuum volume and cold ceramic window volume for specified average input power. Impedance boundary condition of copper at room temperature (300K) was applied to the common surface. Using the built-in macro 'SPARM' and 'HFPOWER', it is possible to calculate the scattering (S) parameters and the total time averaged dielectric losses.

4) By using built-in macro 'ETCHG', HF120 brick element was converted to SOLID90 thermal element. The RF wall losses and the ceramic power losses obtained in Step 3 were applied as thermal heat flux surface loads and heat generation body loads respectively. With the 4 fixed temperature layers (1.8K, 4.2K, 70K and 300K) as external temperature boundary conditions, the 'static + dynamic' temperature profile in the metallic volume can be calculated. For 'static' case, the heat flux surface loads and the heat generation body loads were ignored.

5) With the temperature profile obtained from Step 4 or the previous iteration, new thermal flux surface loads on each of the common surface elements can be recalculated by simple scaling relation ($P \sim \sigma(T)^{-1}$, $P$ is the RF power loss, while $\sigma$ is the electrical conductivity, which is a function of temperature $T$) and reapplied in the following iterating thermal calculation.

The temperature profile along inner and outer conductors has been calculated for different copper coating thickness on warm inner conductor with RRR=10 and 100, supposing the coupler to be operating in continuous regime with 15kW average input power at the 1.3GHz designed frequency.

## 4. Simulation results

### 4.1 RF power losses

Fig. 4 shows one typical RF wall losses distribution for stable travelling wave operation. Here 100μm and 10μm RRR=100 copper were coated on the warm inner and all outer conductors respectively, and the temperature dependence of material properties was also considered. The bellow parts can be clearly identified from the plot.

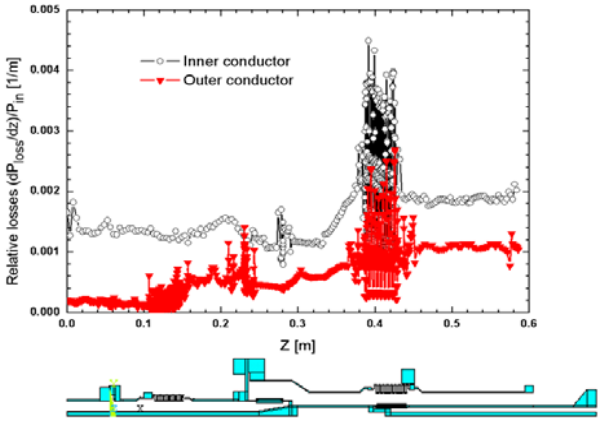

Fig. 4: Typical RF wall losses distribution for TW case

Both S3P [8] and ANSYS were used to calculate the power losses in the 70K ceramic window with ε=9 and tgδ=$10^{-4}$. Fig. 5 shows the electric and magnetic field distribution inside the vacuum and ceramic window volumes. Different from the power loss ratio ($P_{loss,win}/P_{in}$) calculation results $1.94\times10^{-4}$ in Ref. [6], the ratio here is around $1.23\times10^{-4}$ for both FEM codes, similar results were obtained for more finer mesh.

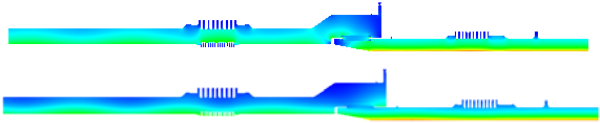

Fig. 5: Electric (upper) and magnetic (lower) field distribution inside the coupler

### 4.2 Thermal calculation results

Fig. 6 and 7 show the temperature distribution along the inner conductor of the TTF-III coupler for RRR=10 and RRR=100. The copper coating on the outer conductor was fixed at 10μm. With the increasing of copper RRR value and coating thickness, maximum temperature rise on the warm inner conductor decreases.

Corresponding to Fig. 6, Fig. 8 shows the temperature distribution along the outer conductor for RRR=10. It can be clearly seen that increasing the copper coating thickness on the warm inner conductor has no big effect on the outer conductor temperature profile. Similar results can be obtained for RRR=100.

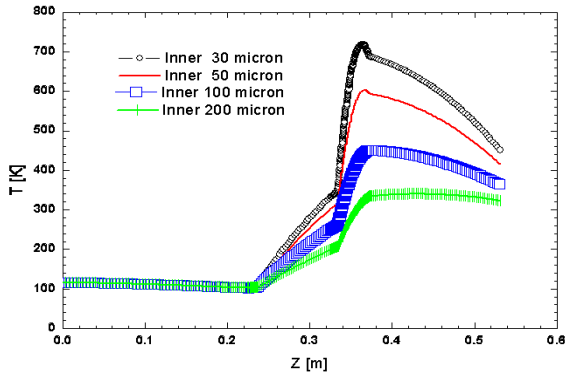

Fig. 6: Inner conductor temperature distribution with 10μm RRR=10 copper coating on outer conductor

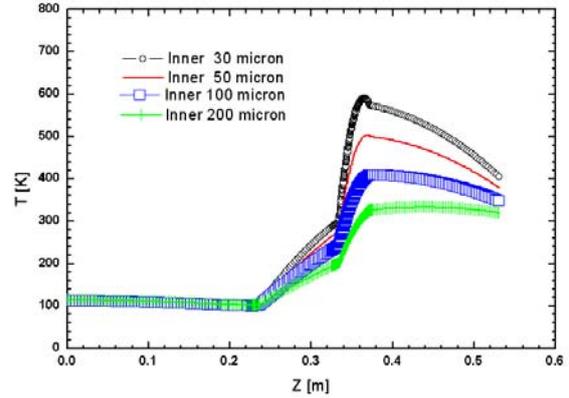

Fig. 7: Inner conductor temperature distribution with 10μm RRR=100 copper coating on outer conductor

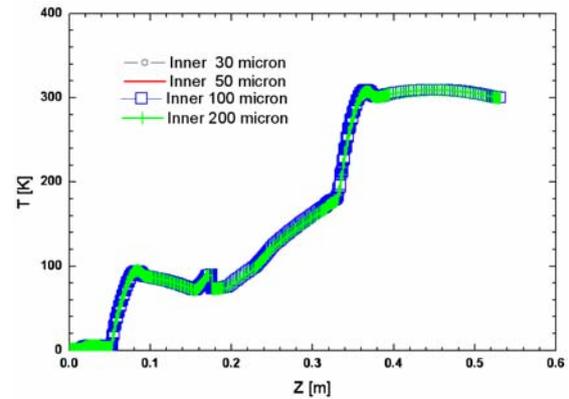

Fig. 8: Outer conductor temperature distribution with 10μm RRR=10 copper coating on outer conductor

Fig. 9 and 10 show the typical temperature distributions for 'static' and 'static + dynamic' cases. 100μm and 10μm copper (RRR=100) was coated on the inner and outer conductors respectively.

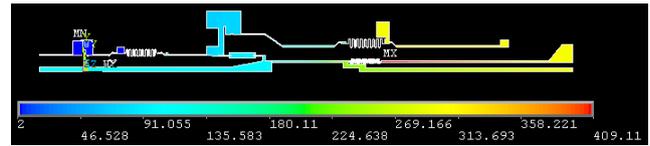

Fig. 9: Typical temperature distribution for 'static' case

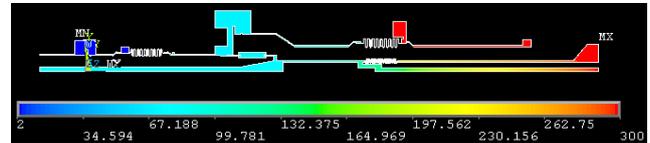

Fig. 10: Typical temperature distribution for 'static + dynamic' case

Table 2 shows all the obtained cryogenic power losses data and the maximum temperature rise on the warm inner conductor. With the increasing of copper coating thickness from 30μm to 200μm, the static cryogenic losses increased from 2.785W to 5.351W for RRR=10 and from 2.859W to 5.580W for RRR=100, the dynamic cryogenic losses decreased 8% for RRR=10 and 1% for RRR=100.

Table 2: Power losses at different temperature layers and the maximum temperature on the warm inner conductor
(copper coating on outer conductor was fixed at 10μm)

| Inner Coating | RRR | Case | P 2K | P 4K | $P_{in}$ 70K | $P_{out}$ | $P_{win}$ | $P_{total}$ | $T_{max}$ [K] |
|---|---|---|---|---|---|---|---|---|---|
| | | | [W] | | | | | | |
| 30 | 10 | Static | 0.008 | 0.240 | 1.454 | 1.083 | 0 | 2.785 | 718 |
| 30 | | Dynamic | 0.357 | 0.790 | 13.45 | 4.574 | 1.84 | 21.01 | |
| 50 | | Static | 0.008 | 0.240 | 1.804 | 1.083 | 0 | 3.135 | 603 |
| 50 | | Dynamic | 0.356 | 0.793 | 13.53 | 4.527 | 1.85 | 21.06 | |
| 100 | | Static | 0.008 | 0.241 | 2.609 | 1.082 | 0 | 3.940 | 450 |
| 100 | | Dynamic | 0.356 | 0.787 | 12.99 | 4.592 | 1.89 | 20.61 | |
| 200 | | Static | 0.008 | 0.241 | 4.021 | 1.081 | 0 | 5.351 | 340 |
| 200 | | Dynamic | 0.361 | 0.792 | 11.86 | 4.515 | 1.81 | 19.34 | |
| 30 | 100 | Static | 0.049 | 0.614 | 1.493 | 0.703 | 0 | 2.859 | 590 |
| 30 | | Dynamic | 0.251 | 0.381 | 12.23 | 4.316 | 1.84 | 19.02 | |
| 50 | | Static | 0.049 | 0.614 | 1.865 | 0.702 | 0 | 3.230 | 503 |
| 50 | | Dynamic | 0.251 | 0.381 | 12.36 | 4.270 | 1.85 | 19.12 | |
| 100 | | Static | 0.049 | 0.614 | 2.719 | 0.702 | 0 | 4.084 | 409 |
| 100 | | Dynamic | 0.251 | 0.381 | 12.55 | 4.341 | 1.89 | 19.41 | |
| 200 | | Static | 0.049 | 0.615 | 4.215 | 0.701 | 0 | 5.580 | 333 |
| 200 | | Dynamic | 0.255 | 0.385 | 12.08 | 4.266 | 1.814 | 18.804 | |

## 5. New cold part design

For coax geometries, the power level for the occurrence of multipacting scales with the 4th power of the diameter of the outer conductor [9]. The multipacting power bands can be increased significantly by using cold part design with larger diameter. It has been shown the TTF-III coupler has a tendency to have long initial high power processing time, which might be caused by multipacting [10]. One new cold part using 60mm coaxial line was designed, which has a relatively longer taper at the pure copper antenna region.

Fig. 11 shows the electric field profile for stable travelling wave operation. Multipacting was simulated with particle tracking code Track3P [8]. Fig. 12 shows the multipacting simulation result. Compared with the results in Ref. [10], the multipacting impact energy has been greatly reduced up to 4MW input power level, indicating the new design will have better power handling capability and less multipacting probability.

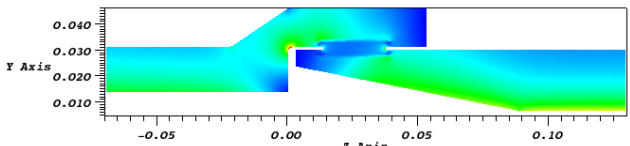

Fig. 11: Electric field profile for the new cold part design

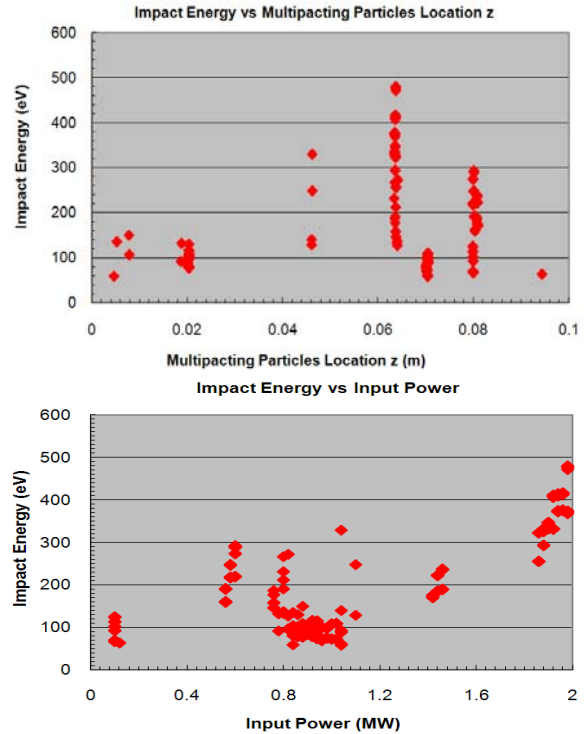

Fig. 12: Multipacting impact energy as a function of axial position (upper) and input RF power level (lower)

## 6. Conclusions

RF-thermal calculations on TTF-III input coupler for copper coating ranging from 30μm to 200μm (RRR=10 and 100) on the warm inner conductor have been done. It shows that the dynamic load is not always constant because of the complicated nonlinear temperature dependence of electric and thermal conductivities. If the tolerable temperature rise is ~150$^o$, copper coating thickness ~100μm would be enough. Increasing copper RRR value does help to reduce the maximum temperature rise. To further decrease the temperature rise on warm inner conductor, air cooled center conductor would significantly help [11].

One new cold coupler part using 60mm coaxial line was designed. From Track3P simulation studies, the new design has less mutipacting probabilities than the old design. The disadvantage of the new design is that the field asymmetry near the antenna region is more severe, which might result in bigger RF kick and wakes and needs to be studied further.